\documentclass{arxiv}
\usepackage{natbib}
\usepackage{multicol}

\usepackage{graphicx}
\usepackage{multirow}
\usepackage{subfigure}
\usepackage{xcolor}
\usepackage{adjustbox}
\usepackage[colorlinks=true,linkcolor=blue,citecolor=blue]{hyperref}
\hypersetup{draft}
\begin{document}

\title{Detection of continuum emission and atomic hydrogen (HI) from comet C/2020 F3 NEOWISE using GMRT}


\author{Sabyasachi Pal\textsuperscript{1,*}, Arijit Manna\textsuperscript{1}}
\affilOne{\textsuperscript{1}Midnapore City College, Kuturia, Bhadutala, Paschim Medinipur, West Bengal, India 721129\\}

\twocolumn[{
\maketitle
\corres{sabya.pal@gmail.com}


\vspace{0.5cm}
\begin{abstract}
Comets are the most primordial objects in our solar system. Comets are icy bodies that release gas and dust when they move close to the Sun. The C/2020 F3 (NEOWISE) was a nearly isotropic comet that was moving in a near-parabolic orbit. The comet C/2020 F3 (NEOWISE) was the brightest comet in the northern hemisphere after comet Hale-Bopp in 1997 and comet McNaught in 2006. In this article, we reported the first interferometric high-resolution detection of the comet C/2020 F3 (NEOWISE) using the Giant Metrewave Radio Telescope (GMRT). We detected the radio continuum emission from comet C/2020 F3 (NEOWISE) with flux density level 2.84($\pm$0.56)--3.89($\pm$0.57) mJy in the frequency range 1050--1450 MHz. We also detected the absorption line of atomic HI with Signal to Noise Ratio (SNR) $\sim$5.7. The statistical column density of the detected HI absorption line was $N_{HI} = (3.46\pm0.60)\times(T_{s}/100)\times10^{21}$ cm$^{-2}$ where we assumed spin temperature $T_{s}$ = 100 K and filling factor $f$ = 1. The significant detection of continuum emission from the comet C/2020 F3 (NEOWISE) in $\sim$21 cm wavelength indicated that it arose from the large Icy Grains Halo (IGH) region.
\end{abstract}

}]
\year{2021}
\setcounter{page}{1}

\section{Introduction}
\label{sec:intro} 
In radio wavelength, the study of the continuum emission of a comet gives information about the thermal emission which originates from large size dust particles and cometary nucleus and the study of the line emission gives the information about the cometary species. In centimeter wavelength, the detection of the radio continuum emission is rare. \citet{Do58} first reported about the detection of radio emission from the comet Arend-Roland 1957 III but the comet C/1973 E1 (Kohoutek) is widely accepted as the first detected comet in radio wavelengths \citep{Ho75}. Later, \citet{Bir74} and \citet{Tur74} detected the OH radical lines at 18 cm from the comet C/1973 E1 (Kohoutek). The size of icy grains of the comet C/1973 E1 (Kohoutek) was in the order of cm. After comet C/1973 E1 (Kohoutek), the OH radical lines were detected from Halley's comet \citep{Cro91, Cro02, Pat91}. The Green bank interferometer detected the radio continuum emission from comet C/1976 (West) with a flux density level of 40 mJy \citep{Hob77}. After the comet C/1976 (West), \citet{dep86} also detected the radio continuum emission from the comet P/Halley using a Very Large Array (VLA) interferometer radio telescope. The Bonn 100 m antenna detected the radio thermal emission from the comet C/1983 H1 (IRAS-ArakiAlcock) at $\lambda$ =  1.3 cm with a continuum level of 9.0$\pm$0.7 mJy \citep{Alt83}. The comet C/1983 H1 (IRAS-ArakiAlcock) was also observed at $\lambda$ = 2 and 6 cm using VLA without detection of any continuum emission \citep{Pat85}. The continuum emission was also detected in mm wavelength from the comet P/Halley using IRAM with flux density 6 mJy at $\lambda$ = 3.5 mm and 52 mJy at $\lambda$ = 1.3 mm \citep{Alt86}. The radio and radar detections of comet P/Halley concluded that the radio continuum emission was arising from the halo of the comet.

\begin{figure*}
	\includegraphics[width=1\textwidth]{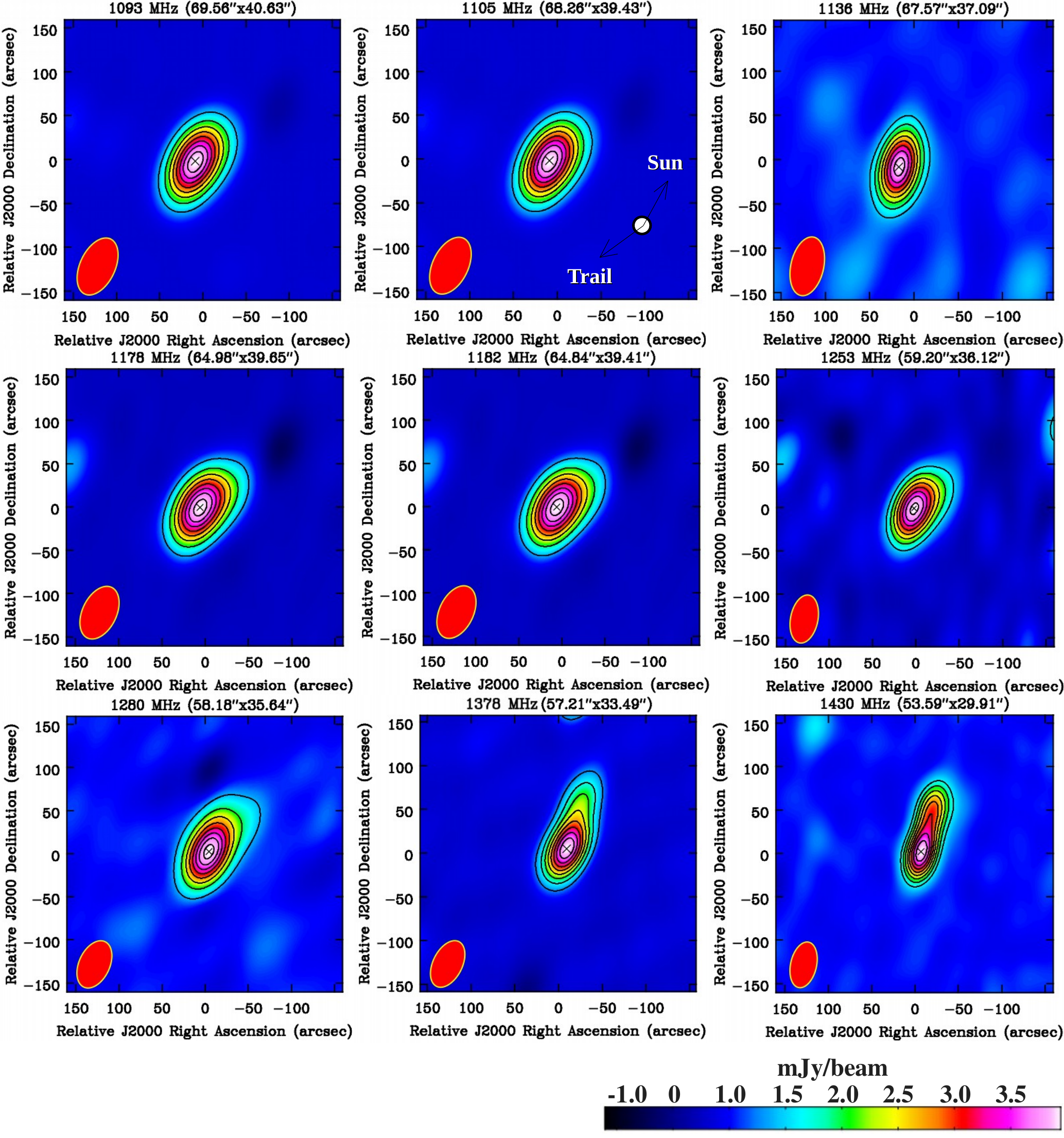}
	\caption{Radio continuum images of comet NEOWISE at frequencies 1093, 1105, 1136, 1178, 1182, 1253, 1280, 1378, and 1430 MHz. The direction of the Sun and the motion of the comet were also shown in the Figure. The red color elliptical shape in the lower-left corner indicates the synthesized beam of the continuum image. The contour levels started at 5$\sigma$ where $\sigma$ = 3 $\mu$Jy/beam was the RMS noise of the continuum image. The contour in each map was increasing by a factor of $\surd$2. In each continuum image, the black cross indicated the comet position provided by the NASA JPL Horizon \citep{gio17}.}
	\label{fig:NEO}
\end{figure*}

The nucleus of comets is usually kilometer-sized with a solid mixture of volatile substances. The volatile components are sublimated and escape from the nucleus of the comet in the form of dust or grains when a comet moves in the direction of the Sun \citep{Whi50}. The nucleus of the comet is surrounded by the large volatile particles \citep{Cam89}. The grains of the comet is visible from Earthbound observer using optical and infrared wavelengths. The grains of the comets are scattered by solar radiation. The chemical composition of cometary grains is very similar to the grains in interstellar medium \citep{Gre82} and circumstellar grains in the disk of young stars \citep{Wei89}. The dust tail of a comet is observed when particle size ($a$) $\ge$ wavelength ($\lambda$) \citep{cro16}.

The water (H$_{2}$O) and hydroxyl (OH) are predominant parents of atomic H which is formed by the photo-dissociation method \citep{kit85}. The core of a comet is surrounded by hydrogen exosphere \citep{Bie64}. In 1968, astronomers confirmed the halo of the comet `Tago-Sato-Kosaka' and `Bennett' which was surrounded by hydrogen in far-ultraviolet wavelength using spacecraft OAO-2 \citep{Cod70, Cod72}. The measurement of intensities of Lyman--alpha and OH resonance by OAO-2 aperture near 309.0 nm wavelength indicated that water was a major species of these comets but it was not predominant.
\begin{table}
	\caption{Orbital parameters of comet NEOWISE}	
	\centering
	\begin{adjustbox}{width=0.5\textwidth}
		\begin{tabular}{|c|c|c|c|c|c|c|c|c|c|}
			\hline 
			Parameters &Symbol&Value\\
			\hline
			Orbit eccentricity& e&0.99918796\\	
			Orbit inclination& i&128.93729229$^{\circ}$\\
			Perihelion distance&q&0.29464931 AU\\
			Aphelion distance&Q&716.6764364031271AU\\
			Semi-major axis&a&358.4855438202229 AU\\
			Orbital period&P&6,912.0000 year\\
			Time of perihelion passage&T&2,459,034.1791 JD\\
			Longitude of ascending node&$\Omega$&61.01410$^{\circ}$\\
			Argument of perihelion&$\omega$&37.27940$^{\circ}$\\
			Longitude of perihelion&L&35.44402$^{\circ}$\\
			Latitude of perihelion&B&28.10517$^{\circ}$\\
			\hline
		\end{tabular}
	\end{adjustbox}\\
	\label{tab:para}
	
\end{table}
Since 1970, many satellites and rockets have studied the hydrogen coma in comets \citep{Mak01}. The Lyman--alpha H line was first detected from bright comet P/Halley at wavelength 121.6 nm by many spacecraft in the northern hemisphere \citep{Mcc86}. 

The comet C/2020 F3 (NEOWISE) (hereafter NEOWISE) is an old nearly isotropic retrograde comet moving in a near-parabolic orbit. It was discovered on March 27, 2020, using the Near-Earth Object program of the Wide-Field Infrared Survey Explorer (NEOWISE) during NEOWISE mission\footnote{{\href{https://neowise.ipac.caltech.edu/}{https://neowise.ipac.caltech.edu/}}}. The comet NEOWISE is one of the brightest comet in the northern hemisphere after the comet Hale-Boppin in 1997 and comet McNaught in 2006 \citep{man21}. The orbit of the comet NEOWISE had a 270 AU inbound semi-major axis and it is projected 255 AU outbound semi-major axis which implies that it is an old long-period comet \citep{com21}. The inbound semi-major axis is the original orbit of the comet and the outbound semi-major axis is the new orbit after the outgassing of the comet near the Sun. Astronomers find the nucleus size of comet NEOWISE was $\sim$5 km using WISE spacecraft by subtracting a fitted dust coma model that means the nucleus size of the comet NEOWISE is slightly larger than that of Rosetta target comet 67P/Churyumov-Gerasimenko, which had a mean diameter of $\sim$4 km \citep{bau20, com21}. The comet NEOWISE makes perihelion on 3 July 2020 at the distance of 0.295 AU from the Sun. The comet was visible by the naked eye for a few weeks near and after the perihelion. Around the comet NEOWISE, the dust emission pattern was observed using optical wavelength \citep{man21}. Recently, the emission lines of atomic OH were detected at frequencies 1612, 1665, 1667, and 1720 MHz using Arecibo Observatory \citep{Sm21}. 
The orbital parameters\footnote{\href{https://ssd.jpl.nasa.gov/horizons.cgi\#results}{https://ssd.jpl.nasa.gov/horizons.cgi\#results}} of comet NEOWISE were shown in Tab.~\ref{tab:para}.

In this article, we presented the first detection of radio continuum emission at the frequency range 1050--1450 MHz and absorption line of HI from comet NEOWISE using Giant Metrewave Radio Telescope (GMRT). In Sect.~\ref{sec:obs}, we briefly described the observations and data reductions. The result and discussion of continuum emission and atomic HI absorption line were presented in Sect.~\ref{sec:Res}. The summary was presented in Sect.~\ref{sec:Dis}.


\section{Observations and data reduction}
\label{sec:obs} 
The comet NEOWISE made its closest approach to Earth on July 23, 2020, 01:14 UT, at a distance of 0.69 AU. We observed the comet NEOWISE on 30th July 2020 (between 08:43--11:34 UTC) using the Giant Meterwave Radio Telescope (GMRT)\footnote{\href{http://www.gmrt.ncra.tifr.res.in/}{http://www.gmrt.ncra.tifr.res.in/}} which is located at Khodad near Pune in India (Latitude: 19$^{\circ}$05$^{\prime}$28.47$^{\prime\prime}$ N, Longitude: 74$^{\circ}$02$^{\prime}$35.44$^{\prime\prime}$ E). This instrument has 30 fully steerable 45 m diameter parabolic dishes. The minimum baseline distance is $\sim$ 100 m and the maximum baseline is $\sim$ 25 km. Currently, GMRT antennas have dual-polarization feed in 150, 230, 327, 610, and 1420 MHz. This radio telescope is running by National Centre for Radio Astrophysics - Tata Institute of Fundamental Research. We used the L-band receiver during the observation of the comet NEOWISE to study the HI spectrum and corresponding continuum emission in the range of 1050--1450 MHz. The L band is the band with the highest observable frequencies amongst the available bands in GMRT which gives the advantage of the better resolution of an image. All available antennas (30) were used during the observation. The number of spectral channels used was 8192 with a spectral resolution of 0.0488 MHz. On the date of observation, the comet was 1.1851 $\times 10^8$ km (0.79 AU) and 1.1405 $\times 10^8$ km (0.76 AU) far from Sun and Earth respectively. The total on-source exposure time was 2.36 hr. We observed the comet with a 15-minute snapshot because the comet was moving very fast. During the observation of the comet NEOWISE, the quasar 3C 286 was taken as a flux and bandpass calibrator while another source 1227+365 was taken as a phase calibrator. 

We used standard calibration using the Common Astronomy Software Application ({\tt CASA}) for initial data reduction and imaging of the comet NEOWISE \citep{mc07}. At first, we imported the interferometric raw data in {\tt CASA} using the task {\tt importgmrt}. For initial data reduction, we flagged the bad data of spectral channels, the first and last records of all the scans, and bad antenna data. The continuum flux density of flux calibrator 3C 286 was scaled with Perley-Butler 2017 with task SETJY \citep{pb17}. The \citet{pb17} flux-density scale is extended downward to $\sim$50 MHz by using recent VLA observations of 20 sources between 220 MHz and 48.1 GHz, and legacy VLA observations at 73.8 MHz. After the absolute flux calibration, we apply the delay and RF bandpass calibration using the flux calibrator. After the initial data calibration, the target source visibilities were smoothed by the spectral resolution 0.0488 MHz for continuum imaging. 
\begin{figure}
	\centering
	\includegraphics[width=0.5\textwidth]{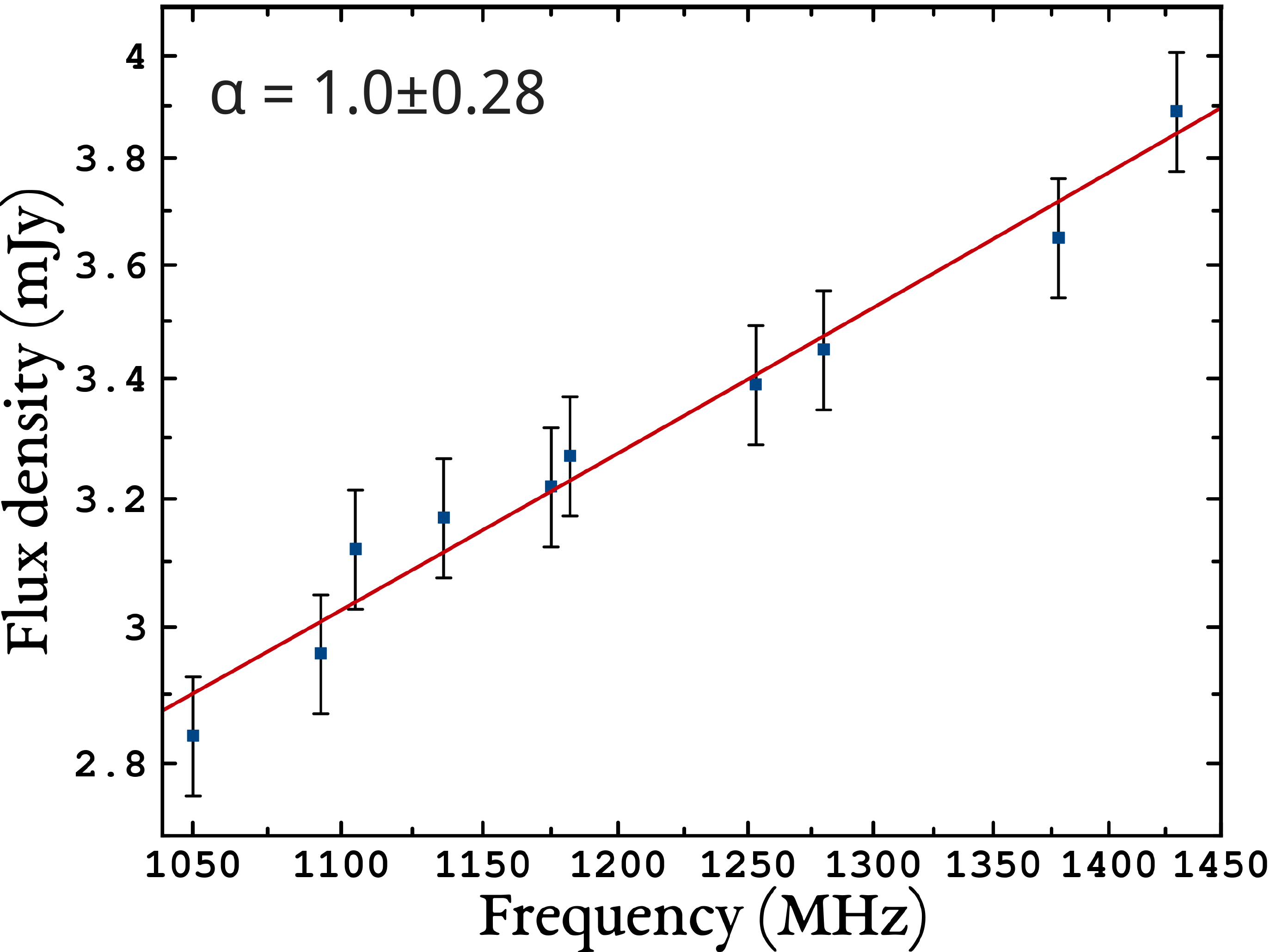}
	\caption{Spectral energy distribution (SED) of comet NEOWISE in the frequency range of 1050--1450 MHz with the variation of flux density in the range of 2.84($\pm$0.56)--3.89($\pm$0.57). The continuum spectrum fitted with a single power law (solid red line) which yielding a global spectral index.}
	\label{fig:spectrum}
\end{figure}
We used task {\tt mstransform} in {\tt CASA} for splitting the target data after gain calibration and transfer of gain calibration to the target. The multi-frequency continuum images of the comet NEOWISE were created by the average of spectral channels (channel bin width 2.0 MHz). After channel average, we used task {\tt tclean} with w-projection \citep{cor08}, and multifrequency synthesis mode with 2nd-order expansion \citep{rau11}. We improve the image sensitivity with a self-calibration procedure using tasks {\tt gaincal} and {\tt applycal}. We used the task {\tt wpbgmrt} for GMRT primary beam correction of continuum image. After the creation of the continuum image of the comet, we used the task {\tt uvcontsub} to subtract the radio continuum emission from the self-calibrated visibility data set. After the continuum subtraction, we created the spectral image of NEOWISE using task {\tt tclean}. We extracted the rotational absorption spectrum of atomic HI against continuum emission from the GMRT spectral data cubes by integrating over the cometary area (disk average) with a $91^{''}$ diameter circular region which centred at RA (J2000) = (12$^{h}$16$^{m}$34$^{s}$.11), Dec (J2000) =  (+30$^\circ$56$^{\prime}$17$^{\prime\prime}$21).

\section{Result and Discussion}
\label{sec:Res} 
\subsection{Continuum emission from the comet NEOWISE}
In Fig.~\ref{fig:NEO}, we have shown the radio continuum images of the comet NEOWISE at frequencies 1093, 1105, 1136, 1178, 1182, 1253, 1280, 1378, and 1430 MHz. The RMS noise level ($\sigma$) of our continuum images was $\sim$3 $\mu$Jy/beam and the synthesized beam size of all images was shown in the header of images.  
The comet was clearly detected in the frequency range of 1050--1450 MHz. 
\begin{figure}
	\centering
	\includegraphics[width=0.5\textwidth]{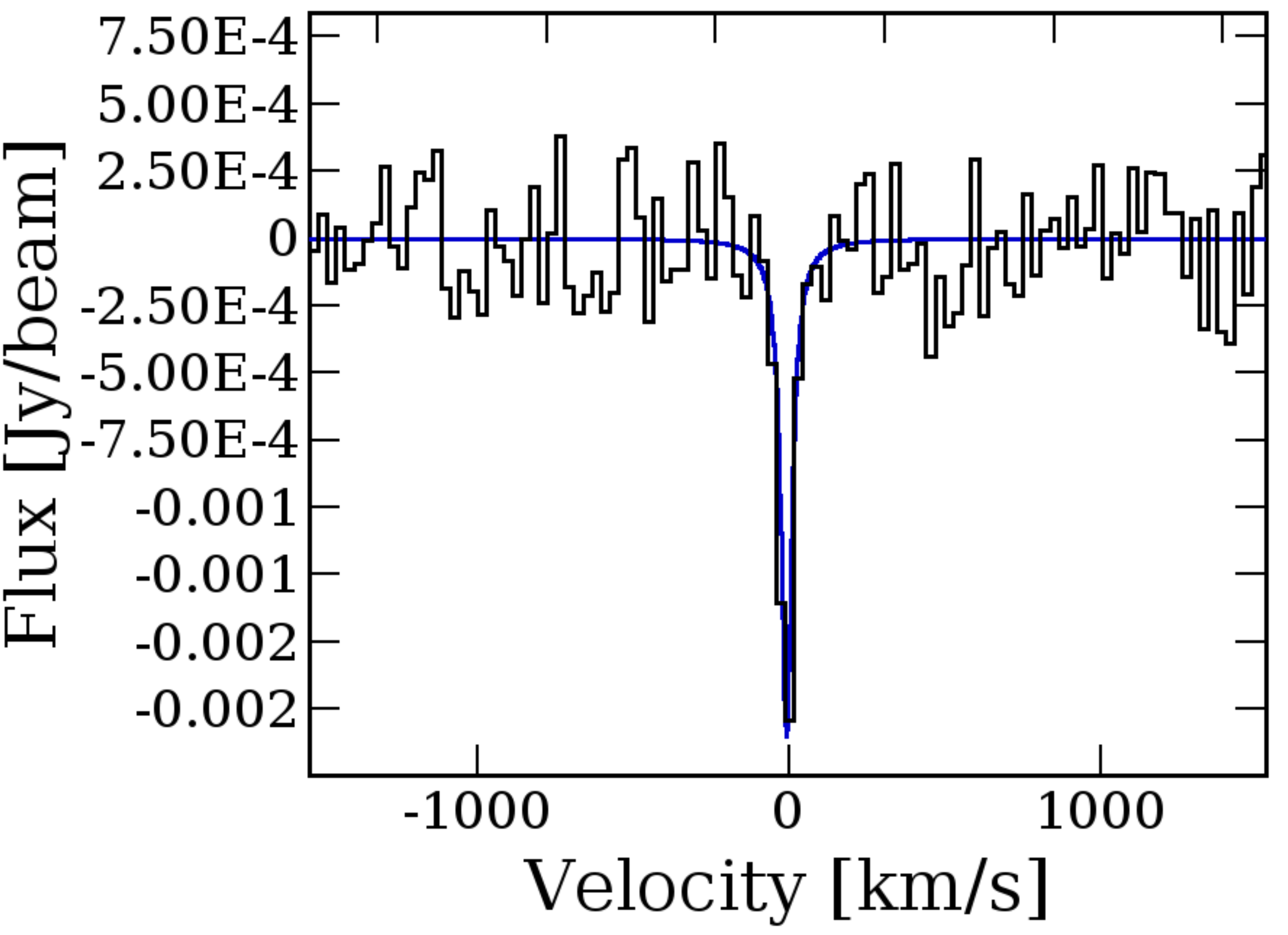}
	\caption{GMRT interferometric spectra of the absorption line of atomic HI from comet NEOWISE which was created from the integrated data cubes with spectral resolution 0.0488 MHz. The synthesized spectrum (blue colour) indicated the single Gaussian model which fitted over the original spectra (black colour) of HI using the MCMC method.}
	\label{fig:absorptionline}
\end{figure}
The flux density of the comet NEOWISE varied in the range of 2.84($\pm$0.56)--3.89($\pm$0.57) mJy which was measured to make a $91^{''}$ diameter circular region in each continuum images which centred at RA (J2000) = (12$^{h}$16$^{m}$34$^{s}$.11), Dec (J2000) = (+30$^\circ$56$^{\prime}$17$^{\prime\prime}$21). The direction of the Sun and the direction of motion of the comet were also shown in Fig.~\ref{fig:NEO}. The elongation of the coma was roughly consistent with the beam pattern, except for 1378 and 1430 MHz, which were noticeably elongated in the Sunward direction, in the opposite direction of the motion of the comet. The elongated continuum emission in the opposite direction of the motion of the comet took place due to ram pressure. The elongation visible in 1378 and 1430 MHz is due to the better resolution of images in these frequencies compared to those of images in lower frequencies. In each continuum image of comet NEOWISE, the black crosshair represented the position of the comet provided by NASA JPL Horizon \citep{gio17}. There was no visible offset between the peak of the radio continuum emission and the location of the core of the comet as expected. The continuum emission from dust particles is significant only at wavelengths smaller than the size of particles \citep{cro16}. Significant detection of comet NEOWISE in $\sim$21 cm indicates the presence of large size of particles in the coma region of the comet. 

The spectral energy distribution (SED) of the comet NEOWISE was shown in Fig.~\ref{fig:spectrum}. The spectral energy distribution was fitted with a single power-law in the frequency range of 1050--1450 MHz and the resultant positive spectral index was $\alpha$ = 1.0$\pm$0.28 (assuming $F_\nu\propto\nu^{\alpha}$). Earlier, \citet{alt99} detected the continuum emission from comet Hyakutake and Hale-Bopp between the frequency range 22 to 860 GHz with the positive spectral index $\alpha$ = 2.8$\pm$0.1 from spectral energy distribution.

\subsection{Detection the absorption line of atomic HI towards the comet NEOWISE}
We detected the absorption line of atomic HI from the comet NEOWISE on 30 July 2020 with a $\sim$5.7$\sigma$ statistical significance level. The absorption line of atomic HI is coming from the hydrogen cloud which surrounded the coma portion of the comet NEOWISE. After the extraction of the absorption spectrum of HI with integrating over the cometary area from GMRT spectral data cubes, we used the Markov Chain Monte Carlo (MCMC) method in {\tt CASSIS} \citep{vas15} for fitting the single Gaussian model over the observed absorption line of atomic HI. Using the fitting of the single Gaussian model over the original transition of atomic HI, the estimated Full width at Half Maximum (FWHM) of the atomic H(I) absorption line was 52.02$\pm$1.94 km s$^{-1}$. Recently, the emission lines of OH were detected from the comet NEOWISE with spectral width $\sim$10 km s$^{-1}$ \citep{Sm21}. The atomic H(I) spectral width was wider than atomic OH lines because the ejection velocity of atomic OH lines from comet nucleus was $\sim$1.05 km s$^{-1}$ and H atom ejection velocity was $\sim$17 times larger than OH atoms due to the momentum conservation i.e $\sim$18 km s$^{-1}$ \citep{Cro89}. So the expected spectral width of atomic H(I) is larger than OH lines. The resultant absorption spectrum of atomic HI with best-fitting Gaussian synthetic spectra was shown in Fig.~\ref{fig:absorptionline}

For a single homogeneous HI absorption line, the statistical column density can be expressed as \citep{ch13}

\begin{equation}
N_ {HI} = 1.823\times10^{18}\times\int T_{s} \tau dv
\end{equation}
where, $T_{s}$ denoted as spin temperature in K, $\tau$ presented the peak optical depth of the absorption spectra, and $\int$ $dv$ is in km s$^{-1}$, which estimated with the integral over the observed absorption line.

We estimated the peak optical depth ($\tau$) of the resultant absorption spectrum was 0.365 and Full Width Half Maxium (FWHM) was 52.02 km s$^{-1}$ which estimated to fit a Gaussian model over the spectrum. So, the corresponding column density of the atomic HI absorption line was $N_{HI} = (3.46\pm0.60)\times(T_{s}/100)\times10^{21}$ cm$^{-2}$ where we assumed the spin temperature ($T_{s}$) 100 K with covering factor ($f$) 1. Recently, \citet{com21} created an image of a hydrogen coma of the comet NEOWISE using SOHO/SWAN on 9 July 2020. We predict that the absorption line of atomic HI is coming from the hydrogen coma of the comet NEOWISE. Before the perihelion, the comet NEOWISE moved very fast in the direction of the Sun, and the frozen volatile component was sublimated and escaped from the nucleus in the form of dust and grains. The outer structure of the coma region is mainly composed of hydrogen gas. Earlier, \citet{kit85} presented the hydrogen coma model using Monte Carlo method which showed the isotropic ejections of H atoms due to the photodissociation of H$_{2}$O (H$_{2}$O + h$\nu$ $\longrightarrow$ H + OH), $V_{H}$ = 20 km sec$^{-1}$) and OH (OH + h$\nu$ $\longrightarrow$ H + O, $V_{H}$ = 8 km sec$^{-1}$).  

\section{Summary}
\label{sec:Dis} 
In this article, we discussed the detection of radio continuum emission in the range of 1050--1450 MHz from the comet NEOWISE which was observed by the GMRT band L receiver on 30 July 2020. The continuum flux density of the comet NEOWSIE varied in the range of 2.84($\pm$0.56)--3.89($\pm$0.57) mJy in the frequency range 1050--1450 MHz. We determined the spectral index $\alpha$ = 1.0$\pm$0.28 (assuming $F_\nu\propto\nu^{\alpha}$) from the spectral energy distribution.

We also detected the absorption line of atomic HI with Signal to Noise Ratio (SNR)$\sim$5.7. Using the peak optical depth and FHWM of the atomic H(I) absorption line, We calculated the statistical column density of atomic HI absorption line was $N_{HI} = (3.46\pm0.60)\times(T_{s}/100)\times10^{21}$ cm$^{-2}$ where assume spin temperature ($T_{s}$) = 100 K with covering factor ($f$) = 1 was used. We measured the flux density and spectral index of the dust continuum emission and derived a HI column density from the atomic HI absorption line. These numbers could be converted into physical parameters of the comet (dust size distribution, HI production rate) but this is beyond the scope of this paper. This present paper summarized the first successful detection of the continuum and HI absorption from comet NEOWISE using GMRT in meter wavelength.


\section*{Acknowledgement}
The plots within this paper and other findings of this study are available from the corresponding author upon reasonable request. We thank the staff of the GMRT for their assistance. We acknowledge Ruta Kale to support during the observation of the comet NEOWISE.
The raw data of NEOWISE reported in this paper are available through the GMRT archive with Proposal Code: ddtC141. This radio telescope is running by National Centre for Radio Astrophysics - Tata Institute of Fundamental Research.

\bibliographystyle{aasjournal}

\end{document}